\documentclass[sn-mathphys]{sn-jnl}% Math and Physical Sciences Reference Style
\jyear{2021}%

\theoremstyle{thmstyleone}%
\theoremstyle{thmstyletwo}%

\theoremstyle{thmstylethree}%

\newcommand{\positron}{e$^+$}
\newcommand{\electronpositron}{e$^\pm$}

\begin{document}

\title[Gamma-ray halos]{Gamma-ray halos around pulsars as the key to understanding cosmic ray transport in the Galaxy}

\author*[1,2]{\fnm{Ruben} \sur{L\'opez-Coto}}\email{rlopezcoto@gmail.com}

\author*[3]{\fnm{Emma} \sur{de O\~na Wilhelmi}}\email{emma.de.ona.wilhelmi@desy.de}

\author[4,5]{\fnm{Felix} \sur{Aharonian}}

\author[6,7]{\fnm{Elena} \sur{Amato}}

\author[4]{\fnm{Jim} \sur{Hinton}}

\affil[1]{\orgname{Istituto Nazionale di Fisica Nucleare, Sezione di Padova}, \orgaddress{\city{Padova}, \postcode{I-35131}, \country{Italy}}}

\affil[2]{\orgname{Instituto de Astrof\'isica de Andaluc\'ia, CSIC}, \orgaddress{\city{Granada}, \postcode{18080}, \country{Spain}}}

\affil[3]{\orgname{DESY}, \orgaddress{\street{Platanenallee 6}, \city{Zeuthen}, \postcode{D-15738}, \country{Germany}}}

\affil[4]{\orgname{Max-Planck-Institut f\"ur Kernphysik}, \orgaddress{\street{P.O. Box 103980}, \city{Heidelberg}, \postcode{D-69029}, \country{Germany}}}

\affil[5]{\orgname{Dublin Institute for Advanced Studies}, \orgaddress{\street{31 Fitzwilliam Place}, \city{Dublin}, \country{Ireland}}}

\affil[6]{\orgname{INAF Osservatorio Astrofisico di Arcetri}, \orgaddress{\street{Largo E. Fermi 5}, \city{Firenze}, \postcode{I-50125}, \country{Italy}}}

\affil[7]{\orgname{Università degli Studi di Firenze}, \orgaddress{\street{Via Sansone 1}, \city{Sesto Fiorentino}, \postcode{I-50019}, \country{Italy}}}



\abstract{Pulsars are factories of relativistic electrons and positrons that propagate away from the pulsar, permeating later our Galaxy. The acceleration and propagation of these particles are a matter of intense debate. In the last few years, we had the opportunity to directly observing the injection of these particles into the interstellar medium through the discovery of  gamma-ray halos around pulsars. This new type of gamma-ray source is produced by electrons and positrons diffusing out of the pulsar wind nebula and scattering ambient photon fields to produce gamma rays. This new field of study comes with a number of observations and constraints at different wavelengths and a variety of theoretical models explaining the characteristics of these halos. We examine the characteristics of the propagation of cosmic rays inferred from the observations of halos and their local and global implications
on particle transport in the Galaxy.
We also discuss the prospects for observations of these sources with facilities such as LHAASO, or CTA or SWGO in the near future.
}

\maketitle



%\linenumbers

%%%%%%%%%%%%%%%%%%%%%%%%%%%%%%%%%
\section{Introduction}
\label{sec:intro}
%%%%%%%%%%%%%%%%%%%%%%%%%%%%%%%%%%

Pulsars, rapidly rotating neutron stars left behind in supernova explosions, have played a critical role in the development of many areas of astrophysics. In non-thermal astrophysics, the nebula around the Crab pulsar was the first known TeV source \citep{Whipple_Crab} and is by far the best-studied site of astrophysical particle acceleration outside the solar system \citep{2008ARA&A..46..127H, 2014RPPh...77f6901B}. The relativistic wind from pulsars is halted at a termination shock and beyond this a synchrotron-radiation-emitting pulsar wind nebula (PWN) forms, apparently dominated by relativistic electron/positron pairs and magnetic fields (see \citep{GaenslerSlane} for a review). The processes of energy conversion from the highly magnetised flow in the vicinity of the pulsar magnetosphere (at $\sim10^6$m) to the particle-dominated one at the parsec-scale nebula are a matter of continuing debate and intense theoretical study (see e.g. \cite{2013MNRAS.431L..48P} and references therein).

 Ideally, in a steady, spherical flow, the gamma-radiation from the confined region should reflect the dominant convective character of propagation (relativistic MHD flow) and the morphology of the nebula - typically elongated/asymmetric relative to the pulsar's position, whereas beyond it a diffusive propagation should dominate. Note however that this paradigm breaks when looking at the emission of the highest energy particles, which, especially in evolved objects, have large Larmor radii, that can be comparable to the system size (more complex MHD simulations and particle transport models are required when looking into PWNe in detail \citep{2003astro.ph..8483R,2017ASSL..446..215D,2016MNRAS.460.4135P}.

\begin{figure}[!ht]
\centering
\includegraphics[width=0.8\linewidth,trim=0 0 326 0, clip]{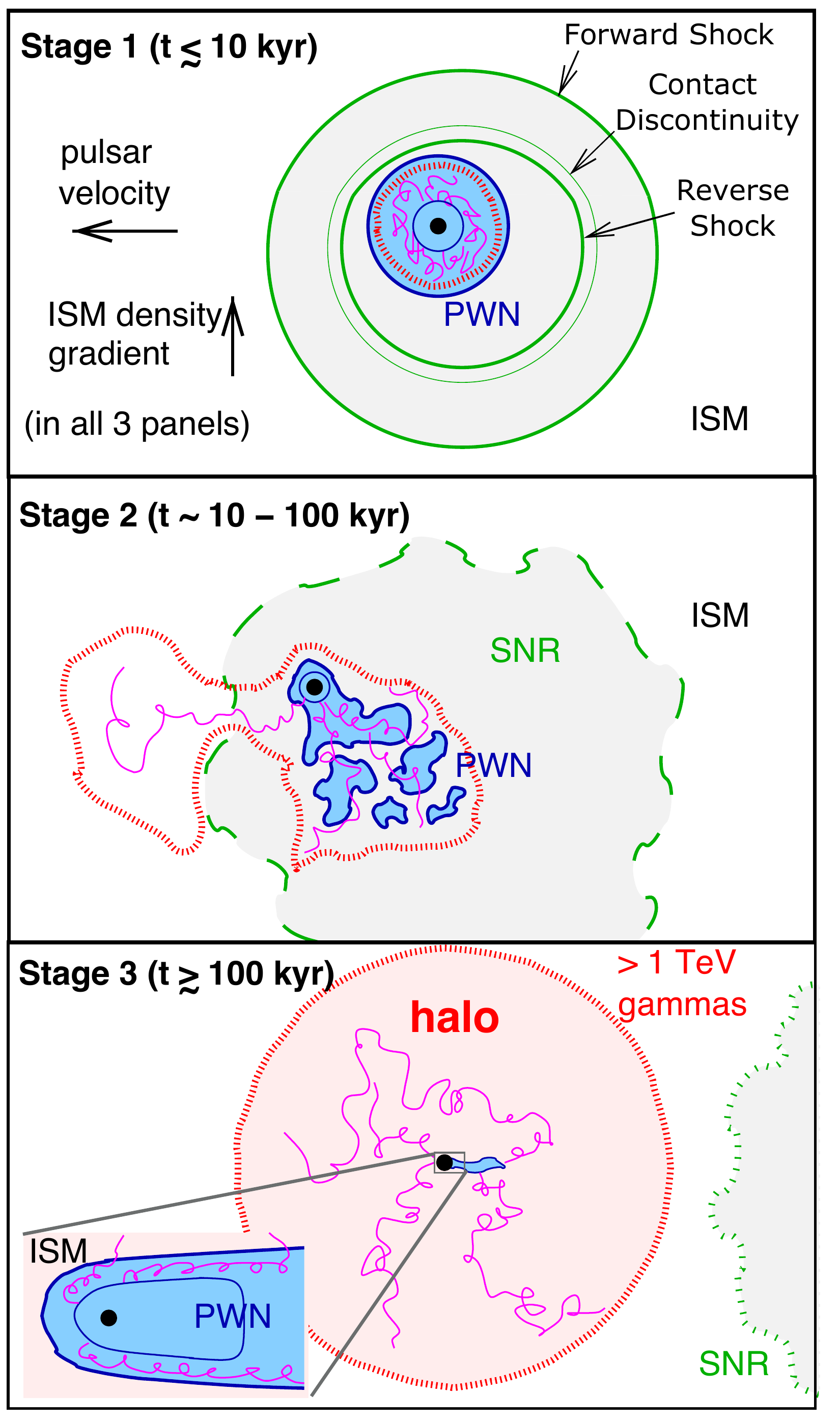}
\caption{Evolutionary stages in the life of a PWN, adapted from Giacinti et al., 2020 \cite{Halo_fraction}, illustrating the early confinement of particles and the later escape of (at least) the higher energy relativistic particles to form a halo visible in TeV gamma rays. %See \cite{Halo_fraction} for details.
}
\label{fig:sketch}
\end{figure}

Figure~\ref{fig:sketch} illustrates the main evolutionary stages of PWNe.
During Stage~1 the PWN is inside the SNR and has not been touched by the reverse shock yet. Stage~2 starts when the PWN is crushed by the reverse shock but the pulsar is still within the SNR: due to the declining magnetic field and the different cooling time of the emitting electrons, X-ray dim, gamma-ray bright 'relic' bubbles can be observed at this stage. Stage~3 starts when the pulsar leaves its parent SNR and high energy particles escape in the ISM: Geminga and PSR B0656+14 \citep{HAWC_Science} are thought to be in this stage. The fringes between Stage 2 and 3 are often unclear,  and there has been considerable discussion of when the term \emph{halo} can legitimately be applied. Here we adopt the term to refer to a population of particles essentially free from their parent PWN, or at least outside of the region in which the nebula is energetically dominant.

Understanding the physics of these halos has implications of fundamental importance for a number of open problems in High Energy Astrophysics. Firstly, halos can be used to probe the evolution of particle acceleration and escape in pulsars and PWNe, which are themselves a unique laboratory for relativistic astrophysics. Secondly, the observation of cooling electrons and positrons traveling freely through the ISM from a well-defined source provides a unique probe of the propagation of relativistic particles.  This
transport is regulated by the diffusion coefficient, which for the ISM has been estimated to be D(10 GeV)$\sim8\times10^{28}$cm$^2$s$^{-1}$ 
\citep{1990acr..book.....B,2011ApJ...729..106T}, through comparison between spallation data and particle transport models. However, this value refers to
the average diffusion coefficient in the ISM, and local variations should be expected. Observations of halos provide information about these variations around powerful electron and positron accelerators. Moreover, these measurements have strong implications on the global electron spectrum.

In this review, we explore the existing experimental constraints on pulsar halos and discuss the theoretical expectations for particle escape and propagation, finally considering the prospects in this area given the powerful new instrumentation on the horizon. 

%%%%%%%%%%%%%%%%%%%%%%%%%%%%%%%%%
\section{Current experimental results}
\label{sec:experimental_results}

The observations of our Galaxy in the TeV energy range with moderate to large field of view sensitive telescopes have revealed a large number of extended regions of gamma rays. In particular, the Galactic plane survey performed by the H.E.S.S. array of Cherenkov telescopes \citep{2005Sci...307.1938A,2006ApJ...636..777A,2018A&A...612A...1H} was the first milestone in the study of such multi-parsec structures, unveiling a large population of PWNe, which dominates the TeV emission of the Galactic plane \citep{2018A&A...612A...2H}. The presence of TeV structures much larger than the X-ray ones was already predicted by \cite{1997MNRAS.291..162A}. The large diversity of TeV PWNe discovered by H.E.S.S. and others IACTs like VERITAS and MAGIC (see e.g \citep{2014A&A...567L...8A,2019ApJ...878..126C}, promoted the classification of PWN evolution in several evolutionary stages, according to the physical properties of the region from which TeV emission originates during the lifetime of a pulsar. This first classification included a free expansion stage, followed by a second phase in which the PWN is interacting with the turbulent plasma left behind by the SN explosion (Stage 1 in Fig. \ref{fig:sketch}); examples of this stage are PWNe like the Crab Nebula or 3C 58. The next stage, also called {\it relic}-stage is one in which a large TeV nebula expands beyond the SNR, altering the surrounding ISM (Stage 2 in Fig. \ref{fig:sketch}).
%Among the latter,
For objects in this stage,
TeV observations with Cherenkov telescopes, combining good energy and angular resolution, allowed disentangling between emission properties in different sub-regions, proving efficient particle cooling and diffusive propagation within the nebula \citep{2020A&A...644A.112H,2019A&A...627A.100H,2012A&A...548A..46H,2020A&A...640A..76P}. 
%\subsection{Large FoV instruments to the rescue} %: HAWC   => Geminga and others
\begin{figure*}[!ht]
\centering
\includegraphics[width=0.95\linewidth]{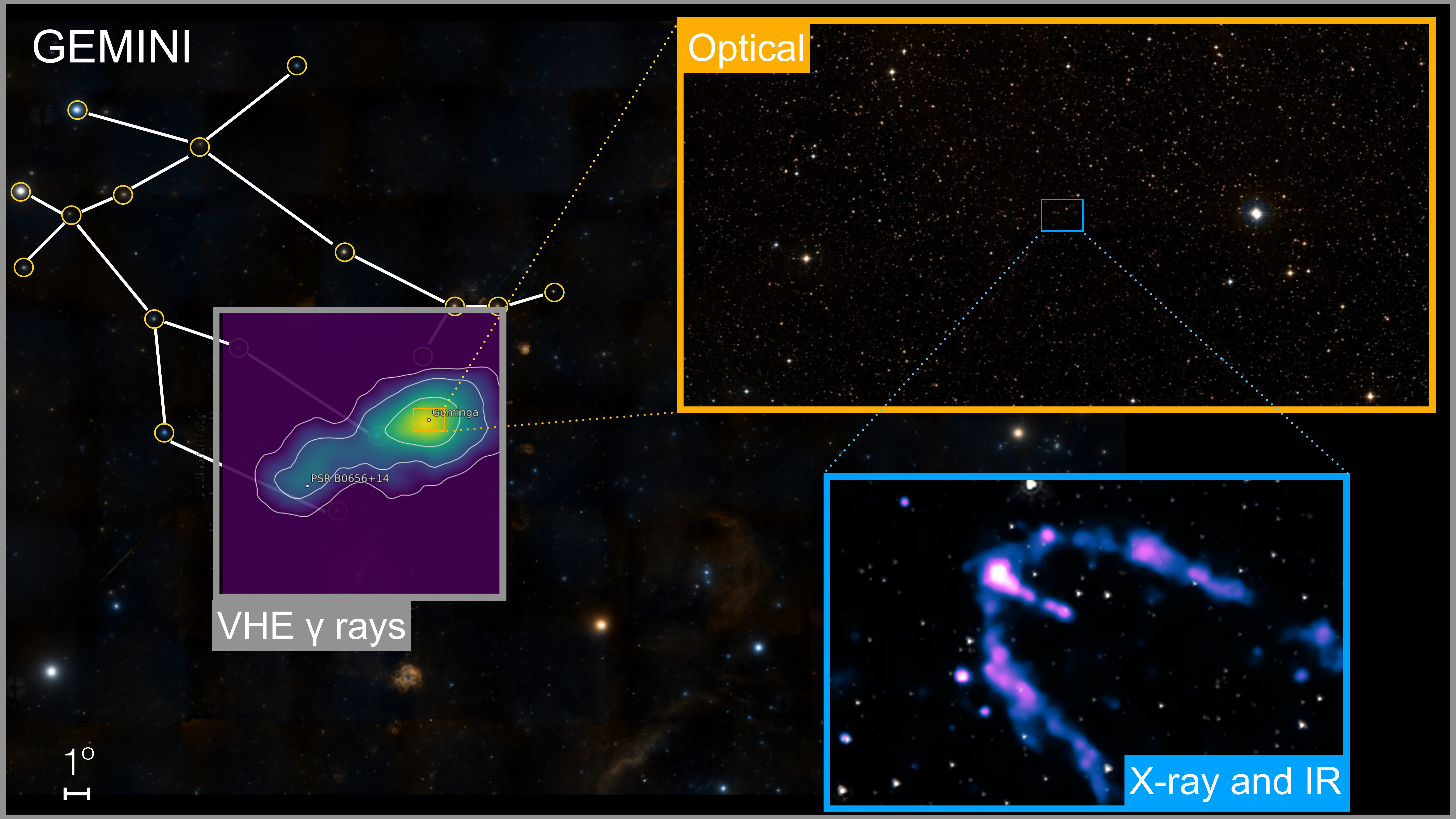}
\caption{Map of the sky region around the HAWC detection of Geminga. VHE gamma-ray inset: Test-statistics (TS) map from the HAWC observations (Abeysekara et al., 2017 \citep{HAWC_Science}). Optical inset: DSS2 obtained using Aladin sky atlas (Bonnarel et al., 2000 \citep{2000A&AS..143...33B}%, 2014ASPC..485..277B}
). X-ray NASA/CXC/PSU (Posselt et al. 2017 \citep{Posselt_2017}); Infrared Map: NASA/JPL-Caltech}
\label{fig:fig2}
\end{figure*}

The advent of large field of view instruments like Milagro \citep{2007ApJ...664L..91A} and currently HAWC and LHAASO in the TeV range \citep{2020ApJ...905...76A,2021Natur.594...33C}, and the {\it Fermi}-LAT satellite in the GeV range \citep{2020ApJS..247...33A}, opened a new window to complete the evolutionary picture of PWNe, 
%being more sensitive 
thanks to improved sensitivity to very large gamma-ray structures (see Fig. \ref{fig:fig2}). The discovery of a large $\sim$5$^{\circ}$ TeV emission around the $\gtrsim100$ kyr old pulsars Geminga %\citep{Milagro_Geminga} 
and PSR B0656+14 triggered a new understanding of the electrons injected within the pulsar environment. The dimension of the halo  ($\sim$25 pc for 100 TeV electrons) has been interpreted as due to slow escape \citep{HAWC_Science} of electrons and positrons accelerated at the 
%shock between the pulsar wind and the ISM.
pulsar wind termination shock.
The most natural explanation for the propagation is that charged particles diffuse in the turbulent magnetic field of the region. Under this assumption, \cite{HAWC_Science} constrained the diffusion coefficient in the region surrounding Geminga and PSR B0656+14 to $D$(100 TeV)$\sim5 \times 10^{27}$ cm$^2s^{-1}$, a value much lower than the average in the ISM, which poses a problem on the origin of this increased level of turbulence with respect to %{\elena{\sout{that obtained using numerical codes to reproduce local CR spectra taking into account propagation over the galactic disk and halo.}} 
that inferred, from local CR spectra, as the average in the Galaxy. Recently, an alternative interpretation to the slow diffusion scenario has been proposed, %{\sout{alternative scenarios to the slow diffusion scenario have been proposed,}}
 in which a combination of ballistic plus diffusive propagation at the same rate as the average in the ISM is used to explain the observed size and TeV emission features \citep{Recchia2021}. The region seen in TeV is characterized by an energy density below that of the ISM, which evidences an outflow of escaping electrons in a region that is not modified by the pulsar itself. The proximity of these pulsars (at $\sim$250~pc), combined with the large field of view makes possible the detection of the otherwise very low TeV gamma-ray surface brightness ( $\sim10^{-12}$TeV cm$^{-2}$s deg$^{-2}$). 
These electrons diffuse away and fill up a large region or {\it halo}, providing a unique clean scenario to study diffusive propagation in the Galaxy. The transition between relic- and halo-stage is blurred and had motivated different classification criteria \citep{Halo_fraction, Linden17}.
An example of this transitional stage is the very extended source HESS\,J1825--137 (21 kyrs) with an unusually large extension of $\gtrsim100$~pc  \citep{2019A&A...621A.116H, 2018ApJ...860...59K}, exceeding the scales anticipated by the standard hydro-dynamical paradigm of PWN formation. The energy-dependence of the morphology strongly suggests advection-dominated transport within the PWN, but the fringes of the emission extend to very large distances and may indicate the presence of unconfined TeV-emitting particles \citep{2019A&A...621A.116H}. Similarly, the 11 kyr-old Vela X shows a compact TeV emission, coincident with a bright X-ray nebula and an extended region with similar spectrum. The extended region also coincides with a radio halo \citep{1958AuJPh..11..550R}, which might be a sign of particles escaping \citep{2006A&A...448L..43A, 2019A&A...627A.100H,2018A&A...617A..78T,2013ApJ...774..110G}. However, the majority of the TeV PWNe are consistent with a relic scenario. These naturally stem from the observational bias towards $\sim$0.2$^{\circ}$ sources at $\sim$5\,kpc distance, for which Cherenkov telescope sensitivity is optimal, in terms of size and flux: at shorter distances, the halo becomes too large to be fully contained in the field of view \citep{HESS_Geminga, 2015arXiv150904224F, 2016A&A...591A.138A}, whereas, for distant sources, the flux might be too faint to be detectable. The good angular resolution permits however a deeper investigation of the morphology in different energy ranges, hinting at transition regions between different propagation regimes. 
Although there are only two firmly identified sources in which emission is produced by already escaped electrons, there have been reports of other halo-candidates \citep{2017ATel10941....1R, 2018ATel12013....1B, 2019ICRC...36..797S, HAWC56TeV, 2021arXiv210609396L} and proposals that this could be the dominant mechanism in several known sources \citep{Hooper_HESS17, 2020PhRvD.101j3035D}. Very recently, the LHAASO collaboration has reported the detection of Geminga \citep{LHAASO_Geminga} and is currently studying it and PSR B0656+14 in detail \citep{Geminga_Monogem_LHAASO}.

The observation of the multi-wavelength counterpart of pulsar halos poses severe challenges. The corresponding size and expected flux are related to the particles cooling regime: in the sub-100 GeV and radio energy band a large emission region with a very hard spectral index is expected, whereas the highest energy electrons are affected by strong cooling, resulting in compact soft X-ray emission (see Fig. \ref{fig:fig2}), as in the case of Geminga \citep{2003Sci...301.1345C, Posselt_2017}. 
Geminga has been investigated in the GeV regime by several authors using data from the {\it Fermi}-LAT telescope. The results are however hampered by the expected number of photons in this energy band: on one hand, the hard uncooled electron spectrum ($\sim$1.8) results in low flux levels below a few tens of GeV; on the other hand, the overwhelming gamma-ray background diffuse emission that dominates the Galactic plane makes it difficult to disentangle the low surface brightness of these halos. Using an energy-dependent, model-dependent template, \citep{2019PhRvD.100l3015D} claimed a detection of the Geminga halo above 10\,GeV%, where a diffusion coefficient of D of [1.6 -- 3.5] $\times 10^{26}$cm$^2$s$^{-1}$ and a size of the halo between 100 and 120 pc. 
. A second analysis by \citep{2019ApJ...878..104X},  using a smaller Region of Interest did not confirm such an extended GeV emission, using a similar physics-motivated template approach (two-zone diffusion spatial templates), although not including the proper motion of the pulsar.% \citep{2019PhRvD.100l3015D} claimed the latest is crucial to reconstruct the gamma-ray emission, whereas other authors dissent \citep{2020arXiv201015731Z}. 
Observations with instruments like CTA or SWGO in the future should provide a clear picture of the evolution in size and spectrum below hundreds of GeV.

In the search for halos, X-ray observations provide different diagnostic tools at different scales. On one hand, the excellent angular resolution of instruments like XMM-{\it Newton} and Chandra provides precise images of pulsars propagating in the ISM, forming a bow-shock which evidences particle escape through the observation of bright filaments \citep{2010MNRAS.408.1216J,  2012ApJ...747...74H, 2014A&A...562A.122P}. On the other, a few-degree soft X-ray halo should emerge, corresponding to the synchrotron emission of electrons powering the gamma-ray source. The detection of such diffuse emission requires a deep observation program involving several pointings. In the case of Geminga, \citep{2019ApJ...875..149L} examined a 1$^{\circ}$ region around the pulsar and obtained an upper limit on the magnetic field below $1 \mu$G, 
%from comparison with the TeV emission, which is expected given the low diffusion coefficient derived from the HAWC observations.
based on X-ray upper limits on the synchrotron emission by the electrons responsible for the emission detected by HAWC. The derived magnetic field is below the mean one in the ISM, pointing to a perturbed medium.

% Back to review the HESS sources looking for hints of halo-like emission => J1825/J1908(?) 
% Lower energies counterpart => Difficulties - LAT / X-rays

%%%%%%%%%%%%%%%%%%%%%%%%%%%%%%%%%
\section{Implications in PWN theory}
\label{sec:theoretical_models}
%%%%%%%%%%%%%%%%%%%%%%%%%%%%%%%%%

The experimental results described above have several implications, regarding not only particle (electrons and positrons) acceleration efficiency, but also their propagation. The maximum particle energy derived from the highest photon energy measured in Geminga ($\approx 300$\, TeV) has strong implication in the acceleration mechanisms: %Let us start discussing the consequences for pulsar physics: from what we have observed so far, 
it appears that %there 
particles
%that the emission comes from particles (electrons and positrons) that 
are accelerated up to a fraction close to 1 of the maximum potential drop available in the pulsar magnetosphere $\Phi_{\rm PSR}=\sqrt{\dot E/c}$ \citep{1969ApJ...157..869G}, with $c$ the speed of light and $\dot E$ the pulsar spin-down power. The bulk of the particles making PWNe bright non-thermal sources are believed to be accelerated at the pulsar wind termination shock (TS). The details of the mechanisms are not clear. The three main proposals are: shock acceleration (see  \cite{2015SSRv..191..519S} for a review), magnetic reconnection \citep{2011ApJ...741...39S} or resonant absorption of ion cyclotron waves \citep{1992ApJ...390..454H, 2006ApJ...653..325A}. All of these mechanisms may in principle reach the required energies (see e.g. \citep{Amato19} for a recent review), but not easily. 
%In fact, it should be noted that $\Phi_{\rm PSR}$ also corresponds to the maximum energy up to which particles can be accelerated at the TS, $E_{\rm max, TS}\approx e R_{\rm TS} B_{\rm TS}$ (where $R_{\rm TS}$ is the TS radius and $B_{\rm TS}$ the local magnetic field strength). 
In fact, the maximum achievable energy at the TS (ignoring all dissipative effects) is determined by the condition that the particle Larmor radius be smaller than the characteristic size of the accelerator ("Hillas criterion"), namely $E_{\rm max,TS}=e R_{\rm TS}\ B_{\rm TS}$, where $R_{\rm TS}$ is the TS radius and $B_{\rm TS}$ the local magnetic field strength. Writing the magnetic pressure at the TS as a fraction $\eta_B$ of th ram pressure of the wind $\dot E/(4 \pi R_{\rm TS}^2 c$), one finds $B_{\rm TS}=(\eta_B^{1/2}/R_{\rm TS}) \sqrt{\dot E/ c}$. From which we derive that the maximum achievable energy at the TS ($\eta_B$=1) is: $E_{\rm max, TS}\approx e \Phi_{\rm PSR}$, {\it i.e.}, in the absence of losses, the maximum energy depends only on the potential drop in the pulsar magnetosphere. Note that this constraint does not depend on the acceleration mechanism. Writing the pulsar spin-down luminosity in units of $10^{36}$ erg/s, we have $E_{\rm max}\approx 1.8 \eta_B^{1/2}\ \dot E_{36}^{1/2}$~PeV (with $\dot E_{36}$ in units of $10^{36}$ erg/s). In particular, for Geminga, with $\dot E_{36}=0.03$, energies of a few hundreds TeV electrons correspond to maximally efficient acceleration.

Once these particles have been accelerated, their escape in the ISM is again a dive into poorly understood physics. The common view of PWNe is that these sources can well be modeled within the framework of relativistic MHD (see e.g. \citep{Olmi16} for a review), where the propagation of particles inside them is governed by advection. This picture is bound to fail at the highest energies and indeed computation of the particle dynamics on top of the e.m. field structure derived from MHD simulations shows that only in a narrow energy range close to $E_{\rm max, TS}$ the fraction of particles that can escape from a PWN becomes sizeable \citep{Olmi19a}. Very interestingly the escaping population is charge-separated, with electrons and positrons escaping in about equal amounts, but along different paths \citep{Olmi19b}. In principle, this would create the conditions for the development, in the PWN vicinity, of a current large enough to have interesting consequences. In fact, one of the proposed explanations for the reduced diffusion coefficient constrained by the HAWC observations is that this results from an enhanced turbulence level produced by the particles escaping the PWN. The well-known resonant streaming instability \citep{1969ApJ...156..445K} does not seem to be effective enough \citep{Evoli18}. The existence of a net current opens the door to the possibility that the fast-growing non-resonant streaming instability \citep{Bell04} can be at work. This requires that the energy density in the current carrying particles is larger than that in the local magnetic field. Such a condition, which is possible in principle to satisfy \citep{Olmi19b}, appears at odds with estimates of the energy density in very high energy electrons derived from modeling of the Geminga halo emission \citep{Halo_fraction}. 
An alternative explanation for the reduced diffusion coefficient is that it results from a local reduction of the magnetic field coherence length, down to $pc$ values, a factor $\approx 10^{-2}$ of what is commonly adopted for MHD turbulence in the Galaxy \citep{2018MNRAS.479.4526L}. In fact, within quasi-linear theory, the diffusion coefficient can be written as
\begin{equation}
    D(E)\approx 2 \times 10^{28} {\rm cm}^2 {\rm s}^{-1}
    \xi_{B,0.1}^{-1}\left(\frac{\lambda_{\rm pc}}{E_{\rm TeV}}\right)^{\alpha-1} B_\mu^{\alpha-2}
%    \left(\frac{\xi_B}{0.1}\right)^\left(\frac{E}{\rm TeV}\right)^{2-\alpha}\left(\frac{B}{\mu G}\right)^{-\alpha+2}\left(\frac{L}{10{\rm pc}}\right)^{alpha-1}
\end{equation}
where $\xi_{B,0.1}=(\delta B/B)^2$ is the ratio between the power of the turbulent and ordered magnetic field normalized to 0.1, $\lambda_{\rm pc}$ is the outer scale of the turbulence in units of pc, $B_\mu$ is the large scale magnetic field in units of $\mu G$ and $E_{\rm TeV}$ is the particle energy in TeV; finally $\alpha$ is the turbulence spectral index with $\alpha=5/3(3/2)$ for a Kolmogorov (Kraichnan) phenomenology. It is clear then that a reduced diffusivity might result from an increased turbulence level (larger $\xi_B$) or a smaller coherence length $\lambda$ of the turbulence. Distinguishing between these two scenarios would only be feasible by looking at particles of higher energies, with Larmor radii comparable to the turbulence coherence length. In that sense, observations with LHAASO should provide crucial information in the understanding of the diffusive regime. Recently, a new solution different from the diffusion-only regime for the propagation of electrons has been proposed. \cite{Recchia2021} studied the halo morphology taking into account that a significant fraction of the propagation of multi-TeV electrons could take place in the ballistic or ballistic-to-diffusive regimes. The ballistic propagation of these electrons in a turbulent magnetic field for distances larger than their Larmor radius could make the gamma-ray morphology of Geminga compatible with the standard diffusion coefficient derived from CR measurements.

%%%%%%%%%%%%%%%%%%%%%%%%%%%%%%%%%
\section{Halos and Galactic Cosmic Rays}
\label{sec:cr_physics}

The escaping particles diffuse further within the Galaxy, ultimately adding up to the sea of Galactic CRs. These CRs can be described as a low-density plasma whose propagation is governed by the diffusion-loss equation \citep{Strong&Moskalenko, Grenier&Strong}. Local measurements of CR fluxes and relative abundances provide insights into the distribution of sources generating them. What pulsar halos directly probe is the transport of electrons and positrons, but since CR propagation in the Galaxy is thought to depend on particle rigidity alone, we expect the propagation of hadrons and leptons to be the same for a given rigidity. Irrespective of the cause of enhanced scattering in the halos, these regions can have important implications for the galactic transport of both leptonic and hadronic CRs, and even put crucial constraints on the origin of the locally measured CR fluxes. 
In this sense, it is important to understand how common pulsar halos are \citep{Sudoh19,Halo_fraction}.

\subsection{Local effects}
Electrons and positrons that escape from these sources can significantly contribute to the local all-electron (\electronpositron) spectrum \citep{Atoyan95}. This provides information about the propagation of these \electronpositron, while the \positron\ fraction (\positron\ flux divided by \electronpositron flux), provides the ratio between the flux of secondary positrons, produced in the collisions of CRs with the ISM, and that of the dominant primary electrons. There is, however, an anomaly known as the ``positron excess'' in the \positron\ fraction, produced by an excess of positrons above the CR induced background \citep{Adriani09, Ackermann12, Aguilar14} above energies of a few GeV that has led to intense speculation on their origin. This excess has been postulated to arise from PWNe \citep{Aharonian95, 2009PhRvL.103e1101Y}, microquasar jets \citep{Gupta14} or dark matter annihilation~\citep{Bergstrom08}.  According to the most commonly accepted propagation theories within the ISM, the highest-energy positrons measured by satellites ($\sim$1~TeV) must originate from a region within $\leq$ few kpc from the Earth ($r_{\rm d}(E) = \sqrt{2 D(E) t_{\rm cool}}(E)$, where $D$(1 TeV)$\sim 10^{29}-10^{30}$ cm$^{-2}$ s$^{-1}$, depending on the assumed energy dependence of the diffusion coefficient, and $t_{\rm cool}$(1 TeV)$\sim300$ kyr, also depending on the assumed energy losses), limiting the number of possible sources behind this phenomenon \citep{Evoli_Amato}.
%According to the most commonly accepted propagation theories within the ISM, the highest-energy positrons measured by satellites ($\sim$1~TeV) must originate from a region within $\leq$500 pc from the Earth ($r_{\rm d}(E) = \sqrt{2 D(E) t_{\rm cool}}(E)$, where $D$(1 TeV)$\sim 10^{29}$ cm$^{-2}$ s$^{-1}$ and $t_{\rm cool}$(1 TeV)$\sim300$ kyr ), limiting the number of possible sources behind this phenomenon \citep{Evoli_Amato}.}

\begin{figure}[!ht]
\centering
\includegraphics[width=1.05\linewidth]{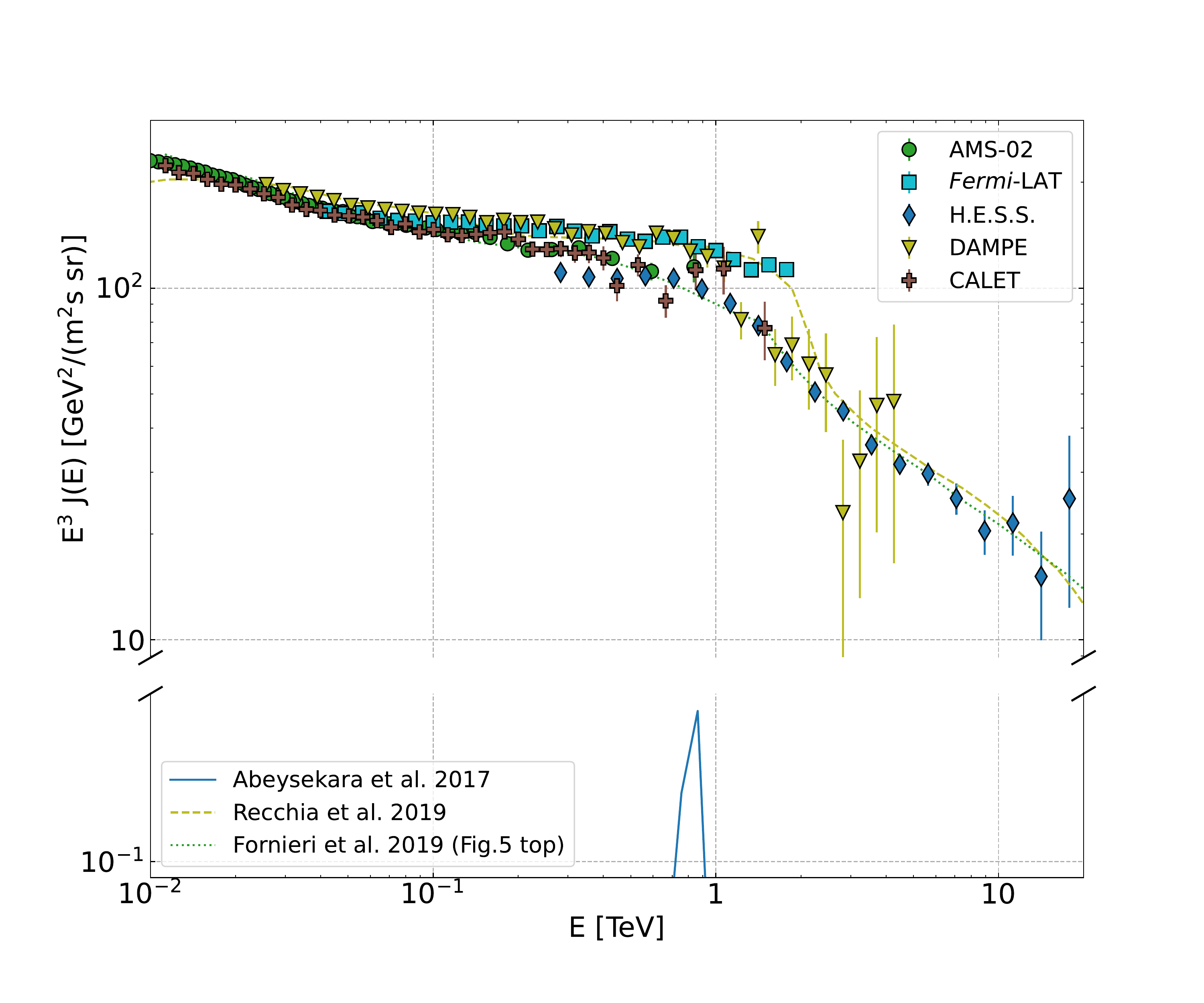}
\caption{Local all-electron spectrum  measured by AMS-02 (Aguilar et al., 2019 \citep{2019PhRvL.122j1101A}), CALET (Adriani et al., 2017 \citep{Adriani17}), DAMPE (Ambrosi et al., 2017 \citep{2017Natur.552...63D}), {\it Fermi}-LAT (Abdollahi et al., 2017 \citep{2017PhRvD..95h2007A}) and H.E.S.S. (Kerszberg et al., 2017 \citep{HESS_electrons}), together with predictions from Reccia et al., 2019 and Fornieri et al., 2020 \citep{2019PhRvD..99j3022R, 2020JCAP...02..009F}, and the flux inferred by Abeysekara et al., 2017 \cite{HAWC_Science}. Errorbars represent 1-sigma statistical errors in the data points.}
\label{fig:all_electron_flux_Earth}
\end{figure}

\begin{figure}[!ht]
\centering
\includegraphics[width=1.05\linewidth]{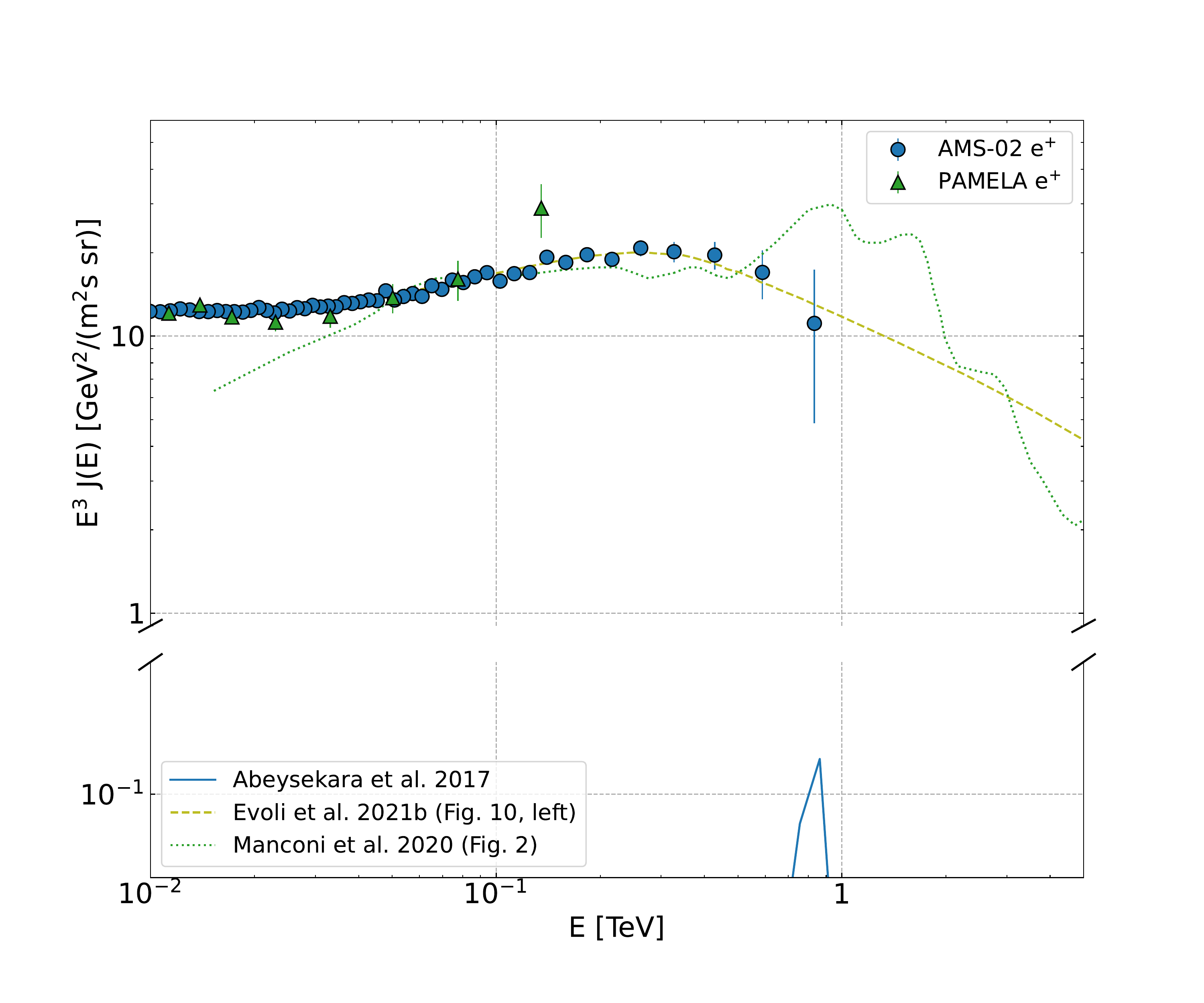}
\caption{Local positron spectrum measured by AMS-02 (Aguilar et al., 2019 \citep{2019PhRvL.122d1102A}) and PAMELA (Adriani et al., 2013 \citep{2013PhRvL.111h1102A}), together with predictions from Evoli et al., 2021 and Manconi et al., 2020 \citep{2021PhRvD.103h3010E, 2020PhRvD.102b3015M}, and the flux inferred by Abeysekara et al., 2017 \cite{HAWC_Science}. Errorbars represent 1-sigma statistical errors in the data points.}
\label{fig:positron_flux_Earth}
\end{figure}

In Fig. \ref{fig:all_electron_flux_Earth} and \ref{fig:positron_flux_Earth}, we can see different estimations from the literature for the local all-electron and positron spectrum, some other recent works on local electron/positron spectra estimations can be found in the literature \citep{2017PhRvD..95f3009L, 2019PhRvD..99d3005L, 2020PhRvL.125e1101E, 2021PhRvD.103h3010E, 2020PhRvD.102b3015M}. The HAWC results on Geminga and PSR\,B0656+14 \citep{HAWC_Science} argued against a significant contribution to the electron and positron spectrum at the Earth by these two pulsars, if assuming a {\it uniform} (one-zone) diffusion coefficient from these two pulsars to the Earth. There has later been extensive literature arguing otherwise \citep{2019ApJ...879...91J, 2020PhRvD.102b3015M, 2017PhRvD..96j3013H, Profumo2018}, by invoking a two-zone diffusion model to describe the propagation of these escaping electrons. In this scenario, the totality of the high-energy electrons and positrons measured at the Earth can be explained using nearby pulsars. 

The recent measurement of the CR all-electron spectrum up to $\sim$20 TeV \citep{HESS_electrons}, in addition to those by DAMPE \citep{2017Natur.552...63D} and CALET \citep{Adriani17} indicates that these electrons and positrons must be generated nearby the Earth because of their cooling due to their interaction with interstellar magnetic and photon fields. Assuming fast diffusion, the origin of these high energy electrons may be the aforementioned pulsars, or sources such as SNRs \citep{2019PhRvD..99j3022R, 2020JCAP...02..009F}. The key to discerning between these two source types is the positron flux, which seems to point to a decrease at the highest energies, disfavoring pulsars as the origin of the highest energy electrons \citep{2019PhRvL.122d1102A}. If on the other hand, we consider a slower diffusion, even an undiscovered pulsar in the Local Bubble, the explanation of the local high energy CR All-Electron Spectrum might still be dominated by pulsars \citep{2018PhRvL.121y1106L}. 
Other observable directly related to the local contribution of a particular source type is the dipole anisotropy, which should in principle pinpoint the origin of the primary accelerator. The current measurements are however still compatible with the different scenarios proposed \citep{Manconi17, 2018PhRvL.121y1106L}.

It remains an open question whether known pulsars \citep{Manconi18}, SNRs \citep{2019PhRvD..99j3022R}, unknown pulsars \citep{2018PhRvL.121y1106L} or DM \citep{Bergstrom08} are still a viable explanation for the local \electronpositron\ flux.

\subsection{Global effects}
The level of magnetic field turbulence with respect to the average in the Milky Way is expected to increase in regions dominated by an active CR accelerator \citep{Fermicocoon}, but it is difficult to find an efficient mechanism that could explain such an increase in the turbulence in the case of halos around pulsars, as discussed in Section \ref{sec:theoretical_models}. Assuming that the existence of pulsar halos implies a slow diffusion coefficient, there are two scenarios in which the existence of pulsar halos can affect the global propagation of CRs in our Galaxy. First of all, we assume that this slow diffusion is only a signature of regions surrounding sources of this type. In this first scenario, if these halos cover a significant fraction of the Galaxy, they could significantly slow down Galactic CR propagation. It is however unlikely that they can impact the overall residence time of hadronic CRs in the Galaxy, which is dominated by propagation in the halo \citep{2018MNRAS.479.4526L, 2017PhRvD..96j3013H}.
The second scenario would be the case that the diffusion coefficient derived from the observation of pulsar halos is a more accurate representation of the average in the Galaxy. This may have profound implications, like a much larger accumulated grammage by CR nuclei (see e.g. \cite{2016PhRvD..94h3003D} for a discussion of these effects around SNRs), which is in any case mostly determined by propagation in the disk, difficult to reconcile with the results obtained with CR propagation codes such as GALPROP \citep{GALPROP}, DRAGON2 \citep{DRAGON2}, PICARD \citep{PICARD} or USINE \citep{USINE}. This constraint must be carefully taken into account by "swiss-cheese" like models of diffusion in the galactic disk \citep{Profumo2018}.

%%%%%%%%%%%%%%%%%%%%%%%%%%%%%%%%%
\section{Conclusions and Prospects}
\label{sec:prospects}
%%%%%%%%%%%%%%%%%%%%%%%%%%%%%%%%%

The "Pulsar - Pulsar Wind - Pulsar Wind Nebula"  concept is a  successful paradigm explaining the link between two major galactic source populations: pulsars, compact relativistic objects, and PWNe, diffuse nonthermal structures filled by magnetic fields and relativistic electrons.  The link is realized through the ultrarelativistic (most likely, cold) electron-positron wind with bulk motion Lorentz factor $\gamma \sim 10^5 - 10^6$. %\cite{ReesGun}. 
The typical magnetic field of almost all PWNe is rather modest - about $10 \mu$G or even less (the Crab Nebula is an atypical  PWN;  its average field of 200-300~$\mu$G is a rare exception - note also that magnetic fields higher than $10 \mu$G have been inferred from modeling efforts, see e.g. \citep{2011MNRAS.410..381B}) but the general trend of low magnetic field holds.  The favorable combination of the low magnetic field and injection of ultrarelativistic electrons at the rate comparable to the pulsar's spin-down luminosity, allows electrons to travel to tens of parsecs from their acceleration sites and form large scale gamma-ray structures which can be detected by the current ground-based instruments. In halo-type emission, we expect spherically symmetric morphology
of gamma rays with radial distribution of electrons close to $1/r$  taken into account the continuous injection of electrons with a constant rate over $\geq 10^4$ years. However, at the highest energies, the radiative losses become an essential factor, and we should see an energy cutoff toward the outskirts of the halo. The model-independent information about the spatial and spectral distribution of electrons provides a unique tool for the extraction of the diffusion coefficient characterizing the propagation of CRs in the galactic disk.  The multi-hundred TeV electrons in the interstellar magnetic fields produce X-ray synchrotron extended sources. The detection of these objects with angular extensions $\geq 1$ degree is challenging but feasible, for bright gamma-ray halos, by eRosita, or planned experiments like AMEGO and AdEPT \citep{2019BAAS...51c.183D}. Thus, the combined X-ray and multi-TeV observations could provide exact independent measurements of the interstellar magnetic field throughout the galactic disk on $\leq 100$~pc scales. Several authors \citep{Linden17, Sudoh19, 2020PhRvD.101j3035D} have computed prospects for the number of halos that could be detected by current and future facilities, using different assumptions. This number ranges from a few, as currently detected, up to hundreds of halos, under the most optimistic assumptions. It is expected that the next-generation detectors, in particular CTA and SWGO, together with the partly completed LHAASO, will dramatically increase the number of identified pulsar halos. Likewise the continuous increment of the already vast {\it Fermi}-LAT dataset will increase the sensitivity to low-surface-brightness sources, not only unveiling more Geminga-like objects but also characterizing the spectrum in the sub-100 GeV energy range.

It would not be an exaggeration to argue that the very task alone of exploration of these standard candles, containing direct information about the energy budget of pulsars in the relativistic electrons, as well as about the CR diffusion coefficient and the magnetic field strength in the interstellar medium, would justify these ambitious ground-based projects.

%%%%%%%%%%%%%%%%%%%%%%%%%%%%%%%%%
\section*{Acknowledgements}
%%%%%%%%%%%%%%%%%%%%%%%%%%%%%%%%%
This article is the result of fruitful discussions during the 1st Workshop on Gamma-ray Halos Around Pulsars (see https://agenda.infn.it/e/GammaHalos). We would first and foremost thank the Scientific Organizing Committee of the Workshop: Alison Mitchell, Stefano Profumo, Diego Torres, Hao Zhou and Roberta Zanin; the Local Organizing Committee: Maria Isabel Bernardos, Alessandro de Angelis, Alessia Spolon and to all the participants: Soheila Abdollahi, 
Amal Abdulrahman, 
Fabio Acero, 
Felix Aharonian, 
Andrea Albert, 
Elena Amato, 
Miguel Araya, 
Thomas Armstrong, 
Vardan Baghmanyan, 
C Baheeja, 
Ali Baktash, 
Yiwei Bao, 
Monica Barnard, 
Ulisses Barres de Almeida, 
Ivana Batkovic, 
Mabel Bernardos Martin, 
Pasquale Blasi, 
Pere Blay, 
Mischa Breuhaus, 
Chad Brisbois, 
Daniel Burgess, 
Senem \capitalcedilla Cabuk, 
Francesca Calore, 
Toms Capistran Rojas, 
Patrizia Caraveo, 
Martina Cardillo, 
Alberto Carrami\~nana, 
Maria Carreon Gonzalez, 
Sabrina Casanova, 
Franca Cassol, 
Safa Chagren, 
Pauline Chambery, 
Tej Chand, 
Songzhan Chen, 
Ji-Gui Cheng, 
Sidika Merve Colak, 
Heide Costantini, 
Roland Crocker, 
Alessandro De Angelis, 
Eduardo de la Fuente Acosta, 
Emma de O\~na Wilhelmi, 
Agnibha De Sarkar, 
Justine Devin, 
Mattia Di Mauro, 
Brenda Dingus, 
Fiorenza Donato, 
Jordan Eagle, 
Christopher Eckner, 
Kathrin Egberts, 
Gabriel Emery, 
Anant Eungwanichayapant, 
Carmelo Evoli, 
Youssef Eweis, 
Kwok Lung Fan, 
Kun Fang, 
Luis Fari\~na, 
Youliang Feng, 
Michele Fiori, 
Henrike Fleischhack, 
Ottavio Fornieri, 
Yves Gallant, 
Gwenael Giacinti, 
Magda Gonzalez, 
Eric Gotthelf, 
Jaziel Goulart Coelho, 
David Green, 
Isabelle Grenier, 
Pietro Grespan, 
Marie-Helene Grondin, 
Yingying Guo, 
Nayantara Gupta, 
Alexander Hahn, 
Hend Hamed, 
Ian Herzog, 
Jim Hinton, 
Bohdan Hnatyk, 
Werner Hofmann, 
Binita Hona, 
Dezhi Huang, 
Zhiqiu Huang, 
Armelle Jardin-Blicq, 
He Jiachun, 
Huang Jiankun, 
Gudlaugur Johannesson, 
Vikas Joshi, 
Francois Kamal Youssef, 
Gottfried Kanbach, 
Dmitry Khangulyan, 
Bruno Khelifi, 
Shota Kisaka, 
Tobias Kleiner, 
Jurgen Kn\"odlseder, 
Dmitriy Kostunin, 
Anu Kundu, 
Michael Kuss, 
Paul Chong Wa Lai, 
Stefan Lalkovski, 
Federico Lavorenti, 
Marianne Lemoine-Goumard, 
Francesco Leone, 
Manuel Linares, 
Tim Linden, 
Ruoyu Liu, 
Sheridan Lloyd, 
Rub\'en L\'opez-Coto, 
Iryna Lypova, 
Kelly Malone, 
Silvia Manconi, 
Vincent Marandon, 
Alexandre Marcowith, 
Jonatan Martin, 
Pierrick Martin, 
Francesco Massaro, 
Razmik Mirzoyan, 
Alison Mitchell, 
Kaya Mori, 
Giovanni Morlino, 
Reshmi Mukherjee, 
Kaori Nakashima, 
Lara Nava, 
Amid Nayerhoda, 
Michael Newbold, 
Melania Nynka, 
Barbara Olmi, 
Elena Orlando, 
Ziwei Ou, 
Maura Pilia, 
Fabio Pintore, 
Illya Plotnikov, 
Troy Porter, 
Raul Ribeiro Prado, 
Elisa Prandini, 
Giacomo Principe, 
Stefano Profumo, 
Habeeb Rahman K. K, 
Brian Reville, 
Chiara Righi, 
Gonzalo Rodr\'iguez Fern\'andez, 
Luca Romanato, 
Bronek Rudak, 
Samar Safi-Harb, 
Takayuki Saito, 
Andres Sandoval, 
Andres Scherer, 
Pooja Sharma, 
Stefano Silvestri, 
Atreyee Sinha, 
Hugh Spackman, 
Alessia Spolon, 
Giulia Stratta, 
Andy Strong, 
Marcel Strzys, 
Takahiro Sudoh, 
Xiaona Sun, 
P.H. Thomas Tam, 
Shuta Tanaka, 
Fabrizio Tavecchio, 
Regis Terrier, 
Luigi Tibaldo, 
Ge Tingting, 
Diego Torres, 
Ramiro Torres Escobedo, 
Michelle Tsirou, 
Naomi Tsuji, 
Antonio Tutone, 
Allard Jan van Marle, 
Juliane van Scherpenberg, 
Gaia Verna, 
Jacco Vink, 
Eda Vurgun, 
Sarah Maria Wagner, 
Jieshuang Wang, 
Xiaojie Wang, 
Jie Xia, 
Gabrijela Zaharijas, 
Silvia Zane, 
Roberta Zanin, 
Davit Zargaryan, 
Darko Zaric, 
Jianli Zhang, 
Yi Zhang, 
Yi Zhang, 
Hao Zhou.

R.L.-C. acknowledges the financial support of the European Union Horizon 2020 research and innovation program under the Marie Sk\l{}odowska-Curie grant agreement No. 754496 - FELLINI. R.L.-C. also acknowledges the financial support from the State Agency for Research of the Spanish MCIU through the ‘Center of Excellence Severo Ochoa’ award to the Instituto de Astrof\'sica de Andaluc\'ia (SEV-2017-0709). EA acknowledges support from ASI-INAF  under grant n.2017-14-H.0) and from INAF under grants “PRIN SKA-CTA”,“INAF Mainstream 2018”and “PRIN-INAF 2019".

\noindent {\bf Author Contributions:} R. L\'opez-Coto and E. de O\~na Wilhelmi coordinated the manuscript writing. All authors meet the journal’s authorship criteria and have reviewed, discussed, and commented on the review content.

\noindent {\bf Competing Interests:} The authors declare that they have no competing interests.

%%%%%%%%%%%%%%%%%%%%%%%%%%%%%%%%%
%\section*{References}
%%%%%%%%%%%%%%%%%%%%%%%%%%%%%%%%%%
%\bibliographystyle{./style/elsarticle-num-names} 
%\bibliographystyle{sn-basic}
%\bibliographystyle{apa}
%\bibliographystyle{./style/mnras} 
\bibliography{references}

\end{document}